\documentclass[aps,prl,reprint,superscriptaddress,longbibliography,floatfix,footnoteinbib]{revtex4-2}
\usepackage{amsmath,amssymb,graphicx,bm}
\usepackage{diagbox}
\usepackage{booktabs}
\usepackage{multirow}
\usepackage{array}
\usepackage{comment}
\usepackage[plainpages=false,pdfpagelabels,colorlinks=true,linkcolor=red,urlcolor=blue,citecolor=blue,pdftitle={Interaction-Split Boundary Spectral Flow in an Integer Hubbard Thouless Pump},pdfauthor={},pdfdisplaydoctitle=true,pdfduplex=DuplexFlipLongEdge]{hyperref}

\begin{document}

\title{Interaction-Split Edge Spectral Flow in an Integer Hubbard Thouless Pump}

\author{Yong-Feng Yang}
\affiliation{Lanzhou Center for Theoretical Physics, Key Laboratory of Quantum Theory and Applications of MoE, Key Laboratory of Theoretical Physics of Gansu Province, and Gansu Provincial Research Center for Basic Disciplines of Quantum Physics, Lanzhou University, Lanzhou, Gansu 730000, China}

\author{Zhao-Rui Tian}
\affiliation{Lanzhou Center for Theoretical Physics, Key Laboratory of Quantum Theory and Applications of MoE, Key Laboratory of Theoretical Physics of Gansu Province, and Gansu Provincial Research Center for Basic Disciplines of Quantum Physics, Lanzhou University, Lanzhou, Gansu 730000, China}

\author{Chen Cheng}
\email{chengchen@lzu.edu.cn}
\affiliation{Lanzhou Center for Theoretical Physics, Key Laboratory of Quantum Theory and Applications of MoE, Key Laboratory of Theoretical Physics of Gansu Province, and Gansu Provincial Research Center for Basic Disciplines of Quantum Physics, Lanzhou University, Lanzhou, Gansu 730000, China}

\author{Hong-Gang Luo}
\email{luohg@lzu.edu.cn}
\affiliation{Institute of Fundamental Physics and Quantum Technology, Ningbo University, Ningbo, 315211 China}
\affiliation{Lanzhou Center for Theoretical Physics, Key Laboratory of Quantum Theory and Applications of MoE, Key Laboratory of Theoretical Physics of Gansu Province, and Gansu Provincial Research Center for Basic Disciplines of Quantum Physics, Lanzhou University, Lanzhou, Gansu 730000, China}

\begin{abstract}
We show that local Hubbard correlations can split the open-boundary spectral flow of an otherwise integer Thouless pump.
Using density-matrix renormalization group calculations for a period-three Hubbard chain with a sliding onsite-repulsion modulation, we identify at filling $\rho=2/3$ a correlated insulating pump with many-body Chern number $C=+2$, corresponding to two units of charge transported per cycle.
A spin-degenerate Hartree/Aubry--Andr\'e--Harper reference model realizes the same integer response through simultaneous edge spectral flow in two independent spin channels.
In the full Hubbard model, however, this integer bulk pump is implemented differently at an open boundary: the edge flow separates into two boundary events, and the intervening low-energy state is a nearly charge-neutral triplet edge excitation.
Thus the quantized bulk charge transport remains integer, but local correlations reorganize its boundary realization into separated charged and neutral spinful sectors.
\end{abstract}

\maketitle

The interplay between topology and strong electronic correlations is a central problem in condensed matter physics~\cite{Colloquium_Hasan2010,Qi2011,Rachel2018}. One-dimensional systems provide an especially sharp setting because charge and spin degrees of freedom can reorganize separately, interaction gaps can replace single-particle band gaps, and boundary spectral flow can be carried by many-body excitations rather than by independent single-particle edge bands~\cite{Luttinger1963,Haldane1981}. Thouless pumps offer a controlled probe of this physics, since in an interacting insulator the charge transported in one adiabatic cycle is fixed by a many-body Chern number~\cite{Thouless1983,Niu1985,Resta1998,Nakajima2016,Lohse2016,Citro2023}. However, this bulk invariant fixes only the total transported charge. It does not specify how the corresponding open-boundary spectral flow is resolved into charge and spin quantum numbers.

This distinction motivates the question addressed here: can a pump with an integer many-body Chern number exhibit an open-boundary spectral flow split by local correlations into distinct quantum-number sectors? Interacting and fractional Thouless pumps have been studied in fermionic, bosonic, and spin systems~\cite{Berg2011,Zeng2016,Taddia2017,Kuno2017,Requist2018,Nakagawa2018,Greschner2020,Jurgensen2023,Jurgensen2025}. Related work has shown that interactions can split two-component Rice--Mele--Hubbard pumping processes~\cite{Bertok2022}, control pumping and its breakdown in quantum-gas experiments~\cite{Walter2023,Viebahn2024}, and enable pumps in nonsliding or nonlinear settings~\cite{Julia2024,Parida2025,Huang2024,Ravets2025,Bai2025,chaudhari2025}. The mechanism studied here is different: the pump is driven by a sliding modulation of the onsite Hubbard repulsion itself. At finite density, this modulation also generates a Hartree channel that can produce a spin-degenerate integer band pump, while the full Hubbard interaction retains local multiplet physics. We show that this local correlation effect reconstructs the open-boundary spectrum: the bulk pump remains an integer-charge pump, but its boundary realization splits into charged edge-transfer events and an intermediate neutral, spinful edge sector.

This should be distinguished from a fractional charge pump. At filling $\rho=2/3$, the many-body Chern number is $C=+2$, corresponding to two units of charge transported per cycle. A Hartree/Aubry--Andr\'e--Harper reference model realizes this response through simultaneous, spin-degenerate edge spectral flow in two independent spin channels. In the full Hubbard model, however, local correlations split this boundary spectral flow into two separated events. Between them, the lowest edge excitation is nearly charge-neutral and carries triplet spin character. This interaction-split boundary spectral flow of an integer Hubbard Thouless pump is the central observation of this Letter.

\textit{Model and diagnostics.---} 
We study a spinful Hubbard chain with a sliding period-three modulation of the onsite interaction, 
\begin{equation} H= -t\sum_{i,\sigma} \left( \hat c^\dagger_{i,\sigma}\hat c_{i+1,\sigma} +\mathrm{H.c.} \right) +\sum_i U_i(\phi)\hat n_{i,\uparrow}\hat n_{i,\downarrow}, 
\label{eq:Hubbard} 
\end{equation} 
where $\sigma=\uparrow,\downarrow$ and $t=1$ is set as the unit of energy. The sliding interaction is 
\begin{equation} 
U_i(\phi)=U\left[1+\delta\cos\left(2\pi i/3+\phi\right)\right], \end{equation} 
where $U$ is the average onsite repulsion and $\delta$ is the modulation strength. Increasing $\phi$ by $2\pi/3$ translates the interaction pattern by one lattice site, while a full $2\pi$ cycle translates it by one spatial period. Unless stated otherwise, we consider chains of length $L=3m$ at filling $\rho=N/L=2/3$, with the ground state in the balanced sector $N_\uparrow=N_\downarrow=N/2=L/3$.

At finite density, the interaction modulation also generates a Hartree channel. To isolate what is beyond this single-particle contribution, we compare Eq.~\eqref{eq:Hubbard} with the spin-degenerate Hartree/Aubry--Andr\'e--Harper (AAH) reference model 
\begin{equation} 
H_{\rm H}(\phi)= -t\sum_{i,\sigma} \left( \hat c^\dagger_{i,\sigma}\hat c_{i+1,\sigma} +\mathrm{H.c.} \right) +\frac{\rho}{2}\sum_i U_i(\phi)\hat n_i, 
\label{eq:Hartree} 
\end{equation} 
with $\hat n_i=\hat n_{i,\uparrow}+\hat n_{i,\downarrow}$. This reference model is not intended as a self-consistent mean-field theory of the Hubbard chain, but rather as a benchmark that retains the leading spin-independent Hartree modulation while removing the local Hubbard multiplet physics. At $\rho=2/3$, it consists of two independent spin copies of a period-three Harper pump~\cite{Harper1955,Aubry1980}, and hence provides the natural spin-degenerate band-pump benchmark.

\begin{figure}[tb] 
\centering 
\includegraphics[width=0.95\columnwidth]{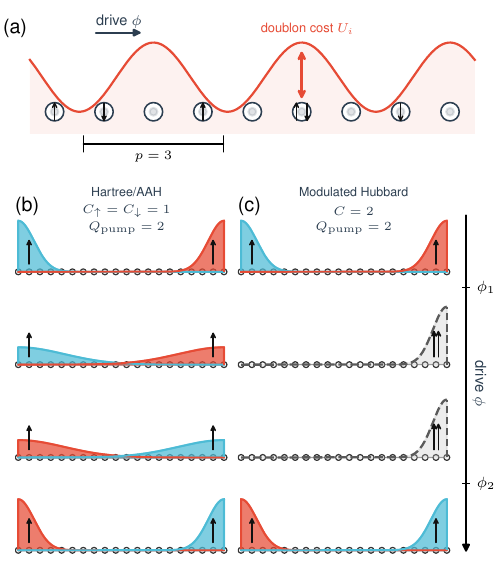}
\caption{ Schematic of interaction-split edge spectral flow. (a) Spinful Hubbard chain at filling $\rho=2/3$ with a period-three sliding onsite interaction [Eq.~\eqref{eq:Hubbard}]. The red profile denotes the modulated onsite repulsion as a doublon cost; advancing $\phi$ translates the interaction pattern. (b) Spin-degenerate Hartree/AAH reference pump. Two independent spin channels exhibit simultaneous edge spectral flow; red and blue packets denote electron- and hole-like edge excitations, while arrows indicate spin. (c) Full modulated Hubbard pump. Local Hubbard correlations split the open-boundary spectral flow into two separated edge events, schematically labeled $\phi_1$ and $\phi_2$. In the intermediate window, the charged boundary response is strongly suppressed, while the remaining low-energy edge mode is a neutral triplet edge excitation, represented by the gray dashed packet. } 
\label{fig:schematic} 
\end{figure}

We obtain ground and low-lying excited states using density-matrix renormalization group (DMRG) calculations~\cite{White1992,White1993,itensor}. Periodic boundary conditions (PBC) are used for bulk gaps, twisted boundary conditions for the many-body Chern number~\cite{Niu1985,Sheng2003,Fukui2005,Xiao2010,Varney2011}, and open boundary conditions (OBC) for edge spectral flow. Charged edge flow is tracked by single-particle addition and removal energies, while the neutral sector is probed by the lowest spin-flip excitation at fixed particle number. Local density, local magnetization, and edge-integrated responses distinguish charged boundary transfer from neutral spin accumulation. Definitions, DMRG convergence checks, finite-size extrapolations, the Hartree reduction, and the Chern-number calculation are given in the Supplemental Material~\cite{SM}.

\textit{Split boundary spectral flow.---} 
Figure~\ref{fig:schematic} summarizes the mechanism. In the Hartree/AAH reference model, each spin component carries Chern number $C_\uparrow=C_\downarrow=1$~\cite{Lang2012prl,Kraus2012a,Kraus2012b}. The total charge pump therefore has $C=+2$, and under OBC the two spin channels exhibit simultaneous edge spectral flow [Fig.~\ref{fig:schematic}(b)]. In the full Hubbard model, the bulk charge transport remains integer, while local Hubbard correlations reorganize the boundary charge and spin sectors and split the open-boundary flow into two separated events [Fig.~\ref{fig:schematic}(c)]. The intermediate sector then contains a neutral triplet edge excitation, with the charged edge response strongly suppressed.

\begin{figure}[tb]
\centering
\includegraphics[width=0.9\columnwidth]{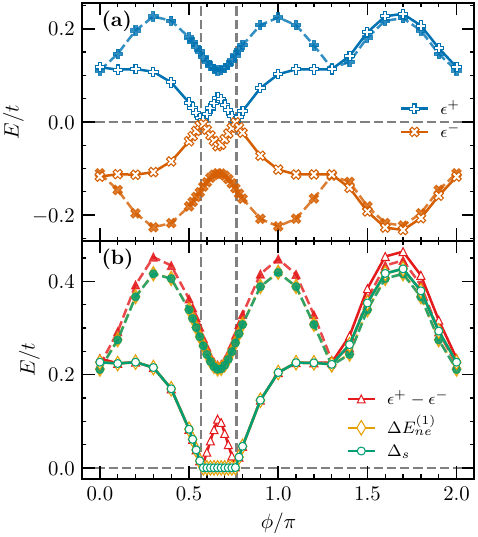}
\caption{ Interaction-split boundary spectral flow at $\rho=2/3$. (a) Particle-addition and particle-removal energies, $\epsilon^{+}$ and $\epsilon^{-}$, measured from a phase-dependent mid-gap reference energy.
(b) Spin-flip gap $\Delta_s$, lowest neutral excitation gap $\Delta E_{\rm ne}^{(1)}$, and fermion charge gap $\epsilon^+-\epsilon^-$. Parameters are $L=72$, $U=2.0$, and $\delta=1.0$. Filled markers and dotted lines denote PBC results, whereas open markers and solid lines denote OBC results. The periodic-chain charge spectrum remains gapped throughout the cycle, whereas the open-chain spectrum shows two boundary events, $\phi_1$ and $\phi_2$. For $\phi_1<\phi<\phi_2$, the open-chain spin-flip gap collapses to zero within numerical resolution while the charge gap remains finite.
}
\label{fig:energy_versus_phi}
\end{figure}

Figure~\ref{fig:energy_versus_phi} verifies this expectation directly in the many-body spectrum for $L=72$ and $U=2.0$. With PBC, the particle-addition and particle-removal spectrum remains gapped throughout the pumping cycle, consistent with a gapped correlated bulk. With open boundaries, in contrast, in-gap states approach the mid-gap reference at two separated phases $\phi_1$ and $\phi_2$ [Fig.~\ref{fig:energy_versus_phi}(a)]. This replaces the single spin-degenerate edge event of the Hartree reference by an interaction-split boundary flow.

The quantum numbers of the intermediate window are resolved by the neutral spectrum. For $\phi_1<\phi<\phi_2$, the open-chain spin-flip gap $\Delta_s$ collapses within numerical resolution, and the lowest neutral excitation $\Delta E_{\rm ne}^{(1)}$ follows the same low-energy structure [Fig.~\ref{fig:energy_versus_phi}(b)]. At the same phases, the fermion charge gap remains finite. Thus the low-energy open-chain degree of freedom is neutral with respect to particle addition or removal but carries spin. This spectral evidence distinguishes the Hubbard pump from a spin-degenerate band pump; the real-space response below shows that the neutral spin excitation is localized near the boundary.

\begin{figure}[tb]
\centering
\includegraphics[width=1\columnwidth]{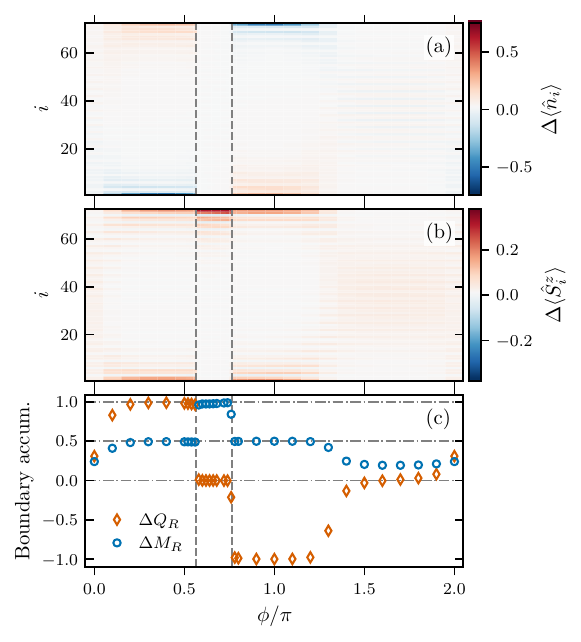}
\caption{ 
Real-space quantum numbers of the split boundary flow. (a) Local charge response $\Delta\langle \hat n_i\rangle$ and (b) local spin response $\Delta\langle \hat S_i^z\rangle$ of the lowest $S^z_{\rm tot}=1$ excitation relative to the balanced ground state at the same particle number. (c) Right-edge integrated charge accumulation $\Delta Q_R$ and magnetization $\Delta M_R^z$, summed over the region $i>2L/3$. Parameters are $L=72$, $U=2.0$, and $\delta=1.0$. Between $\phi_1$ and $\phi_2$, the right-edge charge accumulation nearly vanishes, whereas the right-edge magnetization approaches $\Delta M_R^z\simeq1$, showing that the low-energy edge excitation is nearly charge neutral and spinful. 
}
\label{fig:edge_realspace}
\end{figure}

\textit{Real-space boundary response.---} 
We now examine the spatial structure of the intermediate sector. To remove the static density background, we compute local differences between the lowest spin-flip excited state in the $S^z_{\rm tot}=1$ sector and the balanced ground state at the same total particle number. This construction isolates the real-space response of a fixed-$N$ spin excitation; the precise definitions are given in the Supplemental Material~\cite{SM}.

Figures~\ref{fig:edge_realspace}(a) and \ref{fig:edge_realspace}(b) show the pump-phase evolution of $\Delta\langle \hat n_i\rangle$ and $\Delta\langle \hat S_i^z\rangle$. The charge response is boundary dominated and changes most strongly near the two spectral-flow events identified in Fig.~\ref{fig:energy_versus_phi}. Inside the intermediate window, however, the right-edge charge accumulation is strongly suppressed, whereas the spin response remains sharply localized at the same edge. Although the total charge difference vanishes by construction at fixed $N$, the near-vanishing of the right-edge charge accumulation is nontrivial because the excitation could redistribute charge between the two edges and the bulk. The integrated response in Fig.~\ref{fig:edge_realspace}(c) makes this separation quantitative: $\Delta Q_R\simeq0$ while $\Delta M_R^z\simeq1$ for $\phi_1<\phi<\phi_2$. Together with the near-degeneracy of the OBC spin-flip excitation, this establishes a localized neutral triplet edge excitation generated by the interaction-split spectral flow.

\begin{figure}[!tb]
\centering
\includegraphics[width=1\columnwidth]{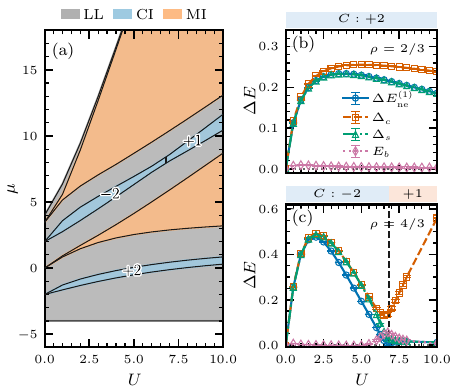}
\caption{
Bulk phase context of the interaction-modulated pump.
(a) Grand-canonical $(U,\mu)$ phase diagram for Eq.~\eqref{eq:Hubbard}, with
$\mu$ the thermodynamic chemical potential.
Colored regions denote gapless Luttinger-liquid (LL), correlated-insulating (CI), and Mott-insulating (MI) regimes.
Labels inside insulating lobes denote their many-body Chern numbers.
(b),(c) Extrapolated bulk gaps versus $U$ at (b) $\rho=2/3$ and (c) $\rho=4/3$ under PBC~\cite{SM}.
The plotted quantities are $\Delta_c$, $\Delta_s$, $\Delta E_{\rm ne}^{(1)}$, and $E_b$.
At $\rho=2/3$, the gapped regime remains in the integer $C=+2$ sector, providing the bulk setting for the boundary spectral-flow splitting discussed above.
The $\rho=4/3$ data are included as a filling-dependent comparison.
}
\label{fig:phase_diagram}
\end{figure}

\textit{Bulk robustness.---} The boundary reconstruction occurs inside a robust integer bulk pump. Figure~\ref{fig:phase_diagram}(a) summarizes the grand-canonical phase map in the $(U,\mu)$ plane, with lobe boundaries extracted from particle-number stability windows as detailed in the Supplemental Material~\cite{SM}. The lobe at $\rho=2/3$ is a correlated insulator with many-body Chern number $C=+2$.

Figure~\ref{fig:phase_diagram}(b) shows the thermodynamic extrapolation of the relevant bulk gaps at $\rho=2/3$. The system is gapless at $U=0$, whereas in the resolved finite-$U$ regime the extrapolated charge and spin gaps are both nonzero, with separated energy scales $\Delta_c>\Delta_s>0$. The many-body Chern number remains $C=+2$ throughout the gapped region studied, and the pair-binding energy is negligible within numerical accuracy. Thus the split boundary spectral flow is not caused by a change in the bulk pumped charge or by bound singlet-pair transport. It is a boundary reconstruction arising from local Hubbard correlations within an otherwise integer-charge pump.

For context, Fig.~\ref{fig:phase_diagram}(c) displays the corresponding bulk gaps at $\rho=4/3$, where stronger coupling produces a distinct filling-dependent topological evolution, including a bulk-gap closing and a change of Chern number. This comparison illustrates that spatially modulated interactions can generate several correlated topological regimes. The central result of this Letter, however, is the $\rho=2/3$ regime: the bulk remains a $C=+2$ integer pump while the OBC spectral flow separates into charged and neutral spin sectors.

\textit{Discussion.---}
We have shown that a sliding modulation of the onsite Hubbard interaction can produce a boundary realization of an integer many-body pump beyond a spin-degenerate band-pump picture.
At filling $\rho=2/3$, the period-three Hubbard chain forms a correlated insulating Thouless pump with many-body Chern number $C=+2$.
A Hartree/AAH reference model realizes this integer response through simultaneous edge flow in two independent spin channels.
The full Hubbard model instead splits the OBC spectral flow into two events and hosts, between them, a localized neutral triplet edge excitation with $\Delta Q_R\simeq0$ and $\Delta M_R^z\simeq1$.
The quantized bulk charge transport therefore remains integer, but local Hubbard correlations change how this integer pump is implemented at an open boundary.

The essential ingredients are a one-dimensional lattice, tunable commensurate filling, and an adiabatically sliding spatial pattern of onsite interactions.
These ingredients are natural for ultracold fermionic atoms in optical lattices, where spatially dependent interactions can be engineered with optical Feshbach techniques~\cite{Clark2015,Yamazaki2010} or modulated cavity fields~\cite{Landig2016,Vaidya2018}.
Recent progress in interacting one-dimensional channels in moir\'e superlattices~\cite{Wang2022,Yang2025,kawakami2026}, transition-metal dichalcogenide arrays~\cite{Jolie2019,Deng2025}, and quasi-one-dimensional van der Waals materials~\cite{Yao2023} also points to solid-state settings where related interaction-reconstructed boundary dynamics may be explored.
Experimentally, the key signatures would be quantized charge pumping, split open-boundary spectral flow, and a neutral triplet response localized near an edge.
More broadly, the results show that an integer many-body Chern number does not uniquely determine the quantum-number structure of the corresponding boundary dynamics.

\begin{acknowledgments}
This research was supported by the National Natural Science Foundation of China (grant Nos. 12174167 and 12247101), the Fundamental Research Funds for the Central Universities (Grant No. lzujbky-2025-jdzx07), and the Natural Science Foundation of Gansu Province (No. 25JRRA799).
\end{acknowledgments}

\textit{Data availability.---}
Data supporting the findings of this study are available from the corresponding author upon reasonable request.

\bibliography{refs}

\clearpage
\onecolumngrid 

\newcommand{\beginsupplement}{%
        \setcounter{table}{0}
        \renewcommand{\thetable}{S\arabic{table}}%
        \setcounter{figure}{0}
        \renewcommand{\thefigure}{S\arabic{figure}}%
        \setcounter{equation}{0}
        \renewcommand{\theequation}{S\arabic{equation}}%
        \setcounter{page}{1}
        \renewcommand{\thepage}{S\arabic{page}}%
        \setcounter{section}{0}
        \renewcommand{\thesection}{S\arabic{section}}%
}

\beginsupplement 

\begin{center}
\textbf{\large Supplemental Material for: Interaction-Split Edge Spectral Flow in an Integer Hubbard Thouless Pump}
\end{center}

\twocolumngrid

\section{S1. DMRG Methodology and Convergence}

We compute the many-body ground state $|\Psi_0(\phi)\rangle$ and low-lying excitations using the density-matrix renormalization group (DMRG) algorithm within the matrix product state (MPS) framework~\cite{Ostlund1995,Dukelsky1998,itensor}. Simulations are performed on finite chains up to $L=96$ sites. We retain a maximum bond dimension of $\chi=3000$ and perform up to 20 sweeps for each value of the pumping phase $\phi$. The discarded weight is typically below $10^{-8}$ in the parameter regimes used for the main-text figures, and convergence is checked by increasing the bond dimension and the number of sweeps.

We use different boundary conditions for complementary diagnostics. Periodic boundary conditions (PBC) are used to extract bulk charge, spin, and neutral excitation gaps, which are then extrapolated to the thermodynamic limit. Twisted boundary conditions (TBC) are used to compute the many-body Chern number. Open boundary conditions (OBC) are used to resolve the boundary spectral flow and the spatial structure of the low-energy boundary response. In the OBC calculations, we monitor the local charge density $\langle \hat{n}_i \rangle = \langle \hat{n}_{i,\uparrow} + \hat{n}_{i,\downarrow} \rangle$ and spin polarization $\langle \hat{S}^z_i \rangle = \frac{1}{2} \langle \hat{n}_{i,\uparrow} - \hat{n}_{i,\downarrow} \rangle$. The corresponding edge-integrated quantities and excitation gaps used in the main text are defined in the next section.

\section{S2. Definition of Energy Gaps and Boundary Observables}

To compare particle-addition and particle-removal spectra at different pump phases under open boundary conditions, we measure all single-particle energies relative to a phase-dependent mid-gap reference energy. This reference is defined as the center of the finite-size charge gap:
\begin{equation}
\mu_0(\phi) = \frac{1}{2}\left[ E_0(N+1, 1/2, \phi) - E_0(N-1, 1/2, \phi) \right],
\label{eq}
\end{equation}
where $E_0(N, S^z, \phi)$ denotes the absolute many-body ground-state energy for a system with $N$ particles and total spin $S^z$ at the pump phase $\phi$.

The single-particle addition energy describes the cost of injecting one electron into the system. We evaluate it by targeting the corresponding $S^z=1/2$ sector. For the $m$-th excitation, this energy is defined as
\begin{equation}
\epsilon^+_m(\phi) = E_m(N+1, 1/2, \phi) - E_0(N, 0, \phi) - \mu_0(\phi),
\label{eq}
\end{equation}
where $E_m$ denotes the energy of the $m$-th state in the target Hilbert-space sector.

Similarly, the single-hole removal energy is defined with respect to the same mid-gap reference:
\begin{equation}
\epsilon^-_m(\phi) = \left[ E_0(N, 0, \phi) - E_m(N-1, 1/2, \phi) \right] - \mu_0(\phi).
\label{eq}
\end{equation}
These definitions reduce finite-size offsets in the addition and removal spectra and allow $\epsilon^+_0(\phi) - \epsilon^-_0(\phi)$ to be used as the finite-size fermion charge gap plotted in the main text.

To construct the global grand-canonical phase diagram, the independent thermodynamic chemical potential $\mu$, which is introduced through $H_{\rm gc}=H-\mu \hat N_{\rm tot}$, determines the stability window of each density plateau. The upper and lower boundaries of the insulating lobes, $\mu^+$ and $\mu^-$, are extracted from the energy cost to add or remove particles. To keep the reference state in the $S=0$ sector and reduce finite-size spin effects, we calculate these boundaries by adding and removing singlet pairs, $\Delta N=\pm2$:
\begin{align}
\mu^+ &= E_0(N+2, 0) - E_0(N, 0) , \\
\mu^- &= E_0(N, 0) - E_0(N-2, 0) .
\end{align}
In the parameter regime emphasized in the main text, the pair-binding energy is negligible within numerical accuracy. The two-particle chemical-potential boundaries therefore provide a stable way to identify the insulating lobes and their thermodynamic charge gaps.

The corresponding bulk charge gap $\Delta_c$, spin gap $\Delta_s$, and pair-binding energy $E_b$ evaluated under periodic boundary conditions are given by:
\begin{align}
\Delta_c &= \frac{1}{2} ( \mu^+ - \mu^- ) \nonumber\\
        & = \frac{1}{2} \left[ E_0(N+2, 0) + E_0(N-2, 0) - 2E_0(N, 0) \right], \\
\Delta_s &= E_0(N, 1) - E_0(N, 0), \\
E_b &= 2E_0(N+1, 1/2) - E_0(N+2, 0) - E_0(N, 0).
\end{align}
These quantities are computed for several system sizes and extrapolated to the thermodynamic limit, $L\to\infty$, using polynomial finite-size scaling.

Finally, to characterize the interaction-reconstructed boundary spectral flow under open boundary conditions, we examine the spatial profile of the lowest spin-flip excitation. The local charge and spin responses are obtained by subtracting the singlet ground-state background from the lowest state in the $S^z_{\text{tot}}=1$ sector:
\begin{align}
    \Delta \langle \hat{n}_i \rangle &= \langle \hat{n}_i \rangle_{S=1} - \langle \hat{n}_i \rangle_{S=0},\\ 
    \Delta \langle \hat{S}^z_i \rangle &= \langle \hat{S}^z_i \rangle_{S=1} - \langle \hat{S}^z_i \rangle_{S=0}.
\end{align}

The integrated right-boundary charge $\Delta Q_R$ and magnetization $\Delta M_R^z$ are obtained by summing these local deviations over the right-edge region used in the main-text analysis:
\begin{align}
    \Delta Q_R = \sum_{i > 2L/3} \Delta \langle \hat{n}_i \rangle,\\
    \Delta M_R^z = \sum_{i > 2L/3} \Delta \langle \hat{S}^z_i \rangle.
\end{align}
These edge-integrated quantities distinguish a charged boundary response from a neutral spin accumulation in the intermediate window between the two split edge events.


\section{S3. Check of the triplet quantum number}

To verify the spin quantum number of the spin-flip excitation, we compute the total-spin expectation value for the open chain,
\begin{equation}
\langle \hat{\mathbf S}_{\rm tot}^2\rangle
= \sum_{i,j=1}^{L}
\left[
\langle \hat S_i^z \hat S_j^z\rangle +
\frac{1}{2}
\left(
\langle \hat S_i^+ \hat S_j^-\rangle
+
\langle \hat S_i^- \hat S_j^+\rangle
\right)
\right],
\label{eq:Stot_direct}
\end{equation}
where \(\hat{\mathbf S}_{\rm tot}=\sum_i\hat{\mathbf S}_i\). The open geometry is used because the excitation of interest is a boundary excitation. We use the same parameters as in the open-chain spectral and real-space calculations: \(L=72\), \(p=3\), \(\rho=2/3\) and \(U=2.0\). The reference sector has \(N_\uparrow=N_\downarrow=L/3=24\), namely \(S^z_{\rm tot}=0\). The spin-flip
sector has \(N_\uparrow=25\) and \(N_\downarrow=23\), giving \(S^z_{\rm tot}=M=1\).

For a state in a fixed \(S^z_{\rm tot}=M\) sector, the spin-algebra identity
\[
\hat{\mathbf S}_{\rm tot}^2
= (\hat S^z_{\rm tot})^2 - \hat S^z_{\rm tot} + \hat S^+_{\rm tot}\hat S^-_{\rm tot}
\]
gives
\begin{align}
\langle \hat{\mathbf S}_{\rm tot}^2\rangle
&= M(M-1) + \langle \hat S^+_{\rm tot}\hat S^-_{\rm tot}\rangle
\nonumber \\
&= M(M-1) + \sum_{i,j=1}^{L} \langle \hat S_i^+\hat S_j^-\rangle .
\label{eq:Stot_fixedM}
\end{align}
The diagonal terms \(i=j\) are included in the double sum. For the spin-flip state,
\(M=1\), so Eq.~\eqref{eq:Stot_fixedM} reduces to
\[
\langle \hat{\mathbf S}_{\rm tot}^2\rangle_{\rm sf}
= \sum_{i,j=1}^{L} \langle \hat S_i^+\hat S_j^-\rangle_{\rm sf}.
\]
A triplet state should satisfy \(\langle \hat{\mathbf S}_{\rm tot}^2\rangle_{\rm sf}\simeq S(S+1)=2\). Since the \(S^z_{\rm tot}=1\) sector excludes singlets, this value identifies the spin-flip excitation as an \(S=1\) triplet rather than a higher-spin state.

\begin{figure}[htb]
    \centering
    \includegraphics[width=0.95\columnwidth]{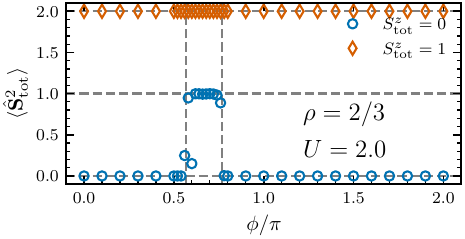}
    \caption{
    Total-spin expectation value $\langle \hat S^2_{\rm tot}\rangle$
    for the open chain at $\rho=2/3$, $U=2.0$, and $L=72$.
    Orange diamonds denote the lowest state in the $S^z_{\rm tot}=1$
    spin-flip sector, while blue circles denote the lowest state found in the $S^z_{\rm tot}=0$ reference sector. The horizontal dashed lines mark the singlet value $\langle \hat S^2_{\rm tot}\rangle=0$ and the triplet value $\langle \hat S^2_{\rm tot}\rangle=2$. The vertical dashed lines indicate the neutral triplet edge interval $\phi_1<\phi<\phi_2$ identified from the open-boundary spectrum and real-space response. The $S^z_{\rm tot}=1$ state remains close to $\langle \hat S^2_{\rm tot}\rangle=2$, confirming its $S=1$ triplet character. The deviations of the $S^z_{\rm tot}=0$ reference state from zero inside the same interval indicate numerical mixing between nearly degenerate singlet and triplet edge states; they are not used to assign the spin quantum number of the spin-flip excitation.
    }
    \label{fig:S_evo}
\end{figure}

Figure~\ref{fig:S_evo} shows \(\langle \hat{\mathbf S}_{\rm tot}^2\rangle\) as a function of \(\phi\). In the \(S^z_{\rm tot}=1\) sector, the result remains close to 2 throughout the pumping cycle. This is expected because the pumping parameter
\(\phi\) does not break spin-rotation symmetry; it changes the energy and spatial character of the excitation, but not its total-spin quantum number. Thus the spin-flip state is an \(S=1\) state for all \(\phi\).

The \(S^z_{\rm tot}=0\) sector is included as a reference. Away from the magnon edge-excitation interval \(\phi_1<\phi<\phi_2\), it gives \(\langle \hat{\mathbf S}_{\rm tot}^2\rangle\simeq0\), consistent with a singlet. Inside \(\phi_1<\phi<\phi_2\), the value deviates from zero. This does not contradict the triplet identification: in the \(S^z_{\rm tot}=0\) sector, both \(|S=0,M=0\rangle\) and \(|S=1,M=0\rangle\) are allowed, and the open-chain edge states can become nearly degenerate. A DMRG calculation that fixes only \(S^z_{\rm tot}\) may therefore converge to a mixture
\[
|\psi\rangle=a|S=0,M=0\rangle+b|S=1,M=0\rangle,
\]
For such a state,
\[
\langle \psi|\hat{\mathbf S}_{\rm tot}^2|\psi\rangle =2|b|^2 .
\]
The \(S^z_{\rm tot}=1\) sector avoids this ambiguity because the singlet component is not present.

The total-spin calculation diagnoses the spin quantum number only. The boundary character is established separately from the real-space profiles and integrated boundary quantities. Therefore, the combined conditions
\[
\langle \hat{\mathbf S}_{\rm tot}^2\rangle_{\rm sf}\simeq2,
\qquad
\Delta Q_R\simeq0,
\qquad
\Delta M_R^z\simeq1,
\]
together with the edge localization of \(\Delta S_i^z\) in \(\phi_1<\phi<\phi_2\), establish a neutral triplet magnon boundary excitation.

In the calculation, a weak spin-dependent pinning field is used only during the initial sweeps to select the desired edge configuration. It is then completely removed, and the final measurements are performed after further convergence with the
unpinned, spin-rotation-symmetric Hamiltonian.

\section{S4. Finite-Size Scaling and the Thermodynamic Limit}

\begin{figure}[htb]
    \centering
    \includegraphics[width=1\columnwidth]{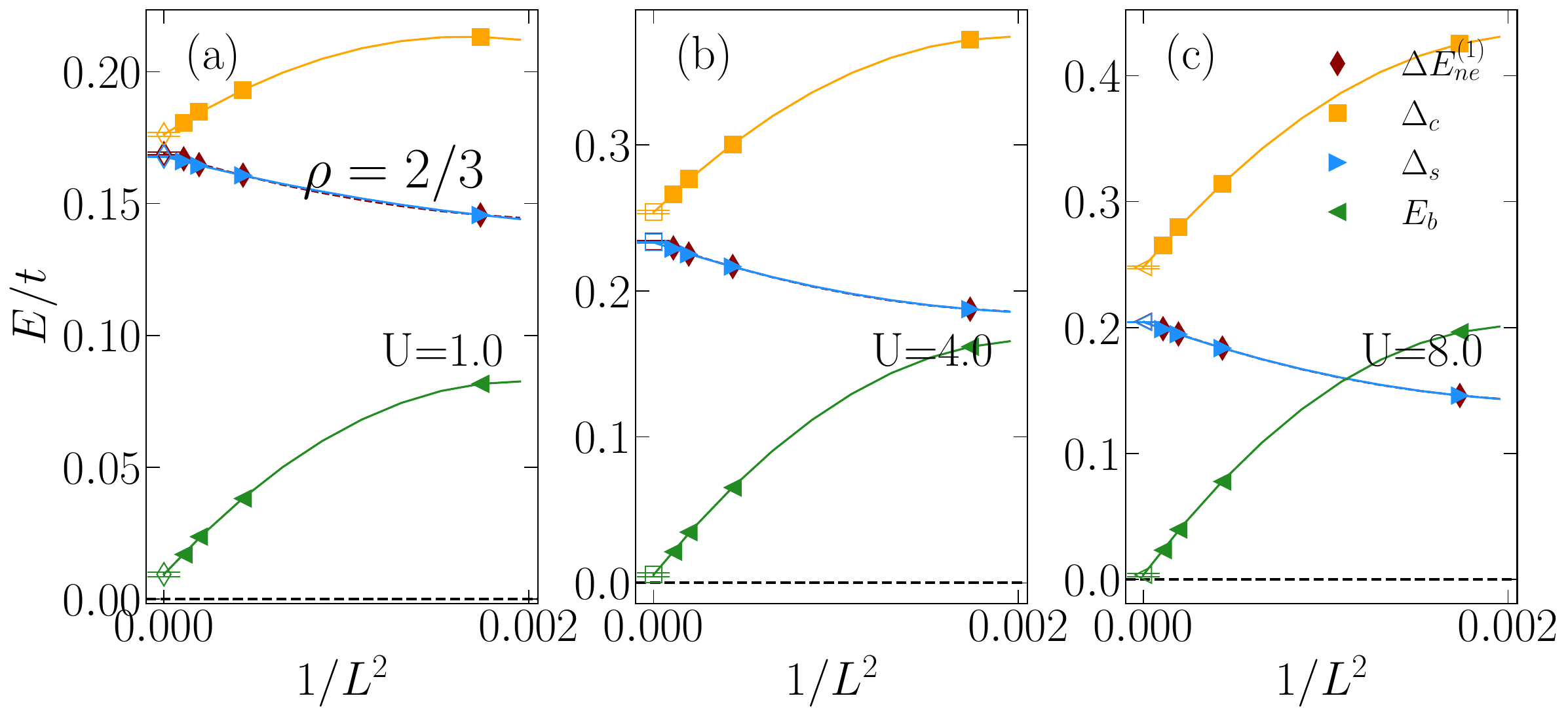}
    \caption{Finite-size scaling of the thermodynamic energy gaps at filling $\rho = 2/3$. The charge gap $\Delta_c$, pair-binding energy $E_b$, spin gap $\Delta_s$, and lowest neutral excitation gap $\Delta E_{ne}^{(1)}$ are plotted versus $1/L^2$ for (a) $U=1.0$, (b) $U=4.0$, and (c) $U=8.0$. Solid symbols denote finite-size DMRG data for $L = 24, 48, 72$, and $96$ under PBC. Open symbols denote the extrapolated values in the thermodynamic limit ($L \to \infty$).}
    \label{fig:fitting_rho2_3}
\end{figure}

\begin{figure}[htb]
    \centering
    \includegraphics[width=1\columnwidth]{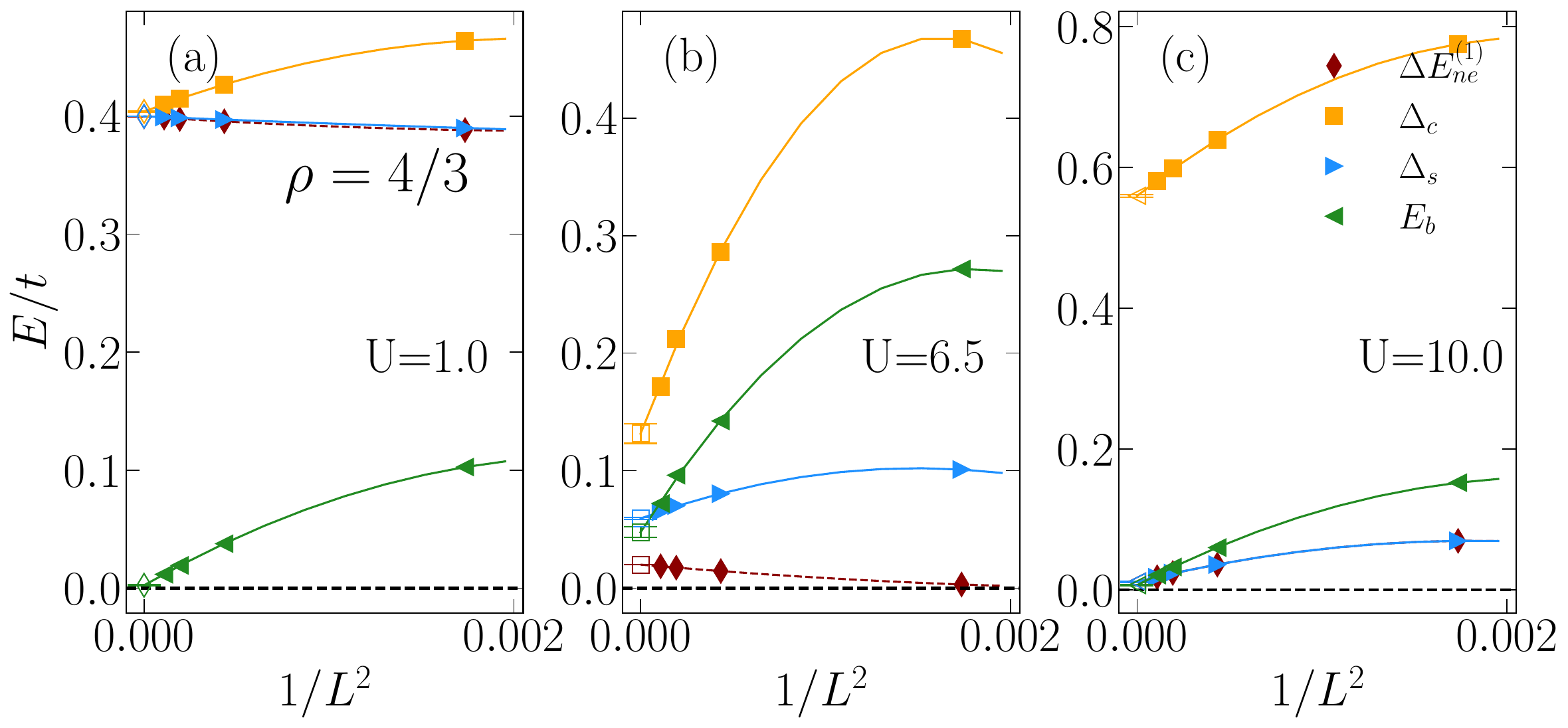}
    \caption{Finite-size scaling of the thermodynamic energy gaps at filling $\rho = 4/3$. The energy gaps are plotted versus $1/L^2$ for (a) $U=1.0$, (b) $U=6.5$, and (c) $U=10.0$. Solid symbols denote finite-size DMRG data for $L = 24, 48, 72$, and $96$ under PBC. Open symbols denote the extrapolated values in the thermodynamic limit ($L \to \infty$).}
    \label{fig:fitting_rho4_3}
\end{figure}

To support the phase diagram presented in Fig.~4 of the main text, we perform a finite-size scaling analysis of the thermodynamic gaps defined in Section S2. The bulk charge gap $\Delta_c$, spin gap $\Delta_s$, lowest neutral excitation gap $\Delta E_{ne}^{(1)}$, and pair-binding energy $E_b$ are evaluated under periodic boundary conditions for system sizes $L = 24, 48, 72$, and $96$. The thermodynamic limit ($L \to \infty$) is obtained by fitting the finite-size data to a polynomial scaling form. For the system sizes considered here, the leading size dependence is well described by plotting the data against $1/L^2$.

Figure~\ref{fig:fitting_rho2_3} shows the scaling results for the main filling $\rho=2/3$. The extrapolated charge gap $\Delta_c$ remains finite for the interaction strengths shown, supporting the existence of a correlated insulating bulk. The spin gap $\Delta_s$ is also finite but remains smaller than the charge gap, giving a separated hierarchy of energy scales, $\Delta_c>\Delta_s>0$. This hierarchy is consistent with the interpretation that the insulating state is not simply a spin-degenerate single-particle band insulator. The pair-binding energy $E_b$ is negligible within the numerical resolution of the extrapolation, indicating that the $C=+2$ pump discussed in the main text is not dominated by bound singlet-pair transport. The lowest neutral excitation gap $\Delta E_{ne}^{(1)}$ tracks the spin gap within the extrapolation uncertainty, supporting the identification of the lowest neutral bulk excitation as predominantly spin-like.

\begin{figure}[htb]
    \centering
    \includegraphics[width=1\columnwidth]{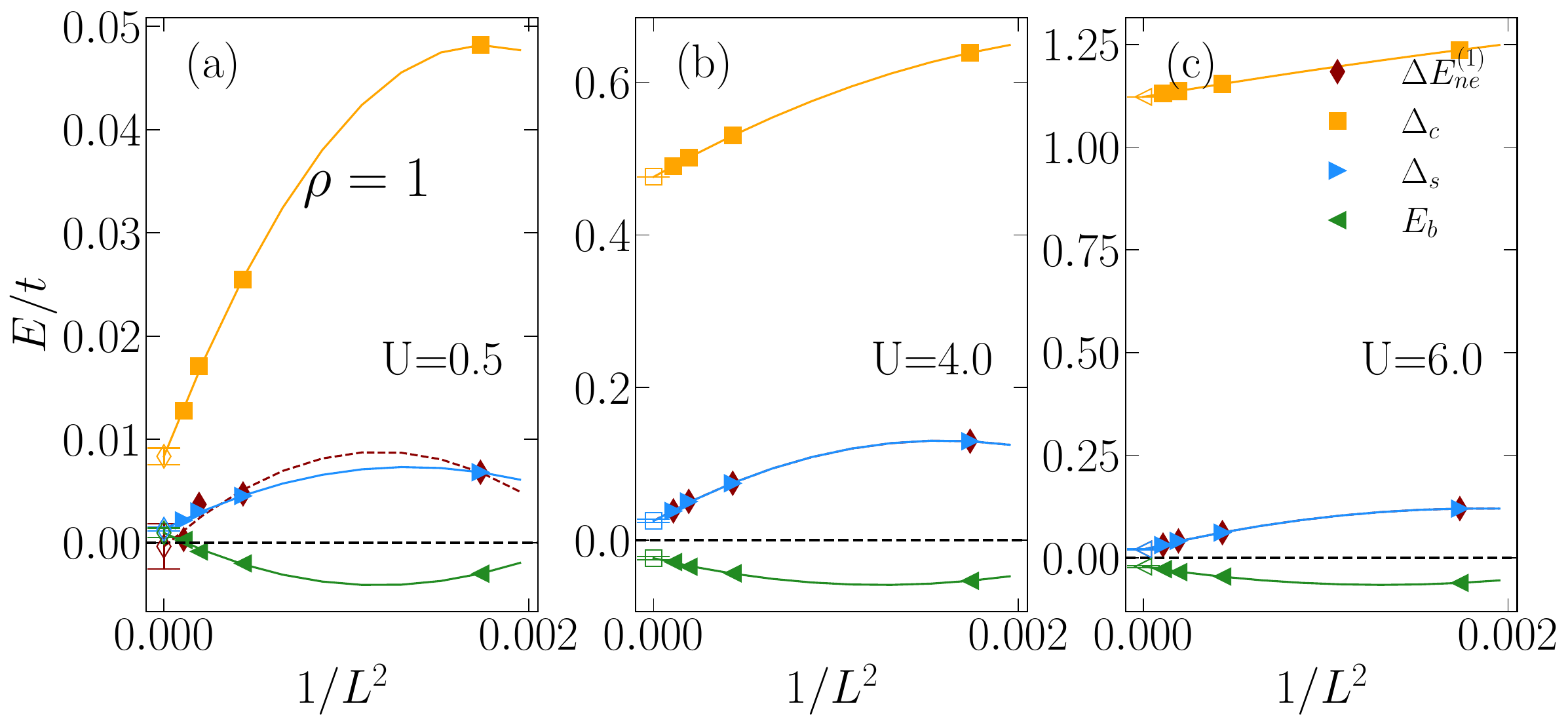}
    \caption{Finite-size scaling of the thermodynamic energy gaps at filling $\rho = 1$. The energy gaps are plotted versus $1/L^2$ for (a) $U=0.5$, (b) $U=4.0$ and (c) $U=6.0$. Solid symbols denote the exact DMRG data for finite-size systems ($L = 24, 48, 72$, and $96$ under PBC). Open symbols denote the extrapolated values in the thermodynamic limit ($L \to \infty$).}
    \label{fig:fitting_rho1}
\end{figure}

\begin{figure}[htb]
    \centering
    \includegraphics[width=1\columnwidth]{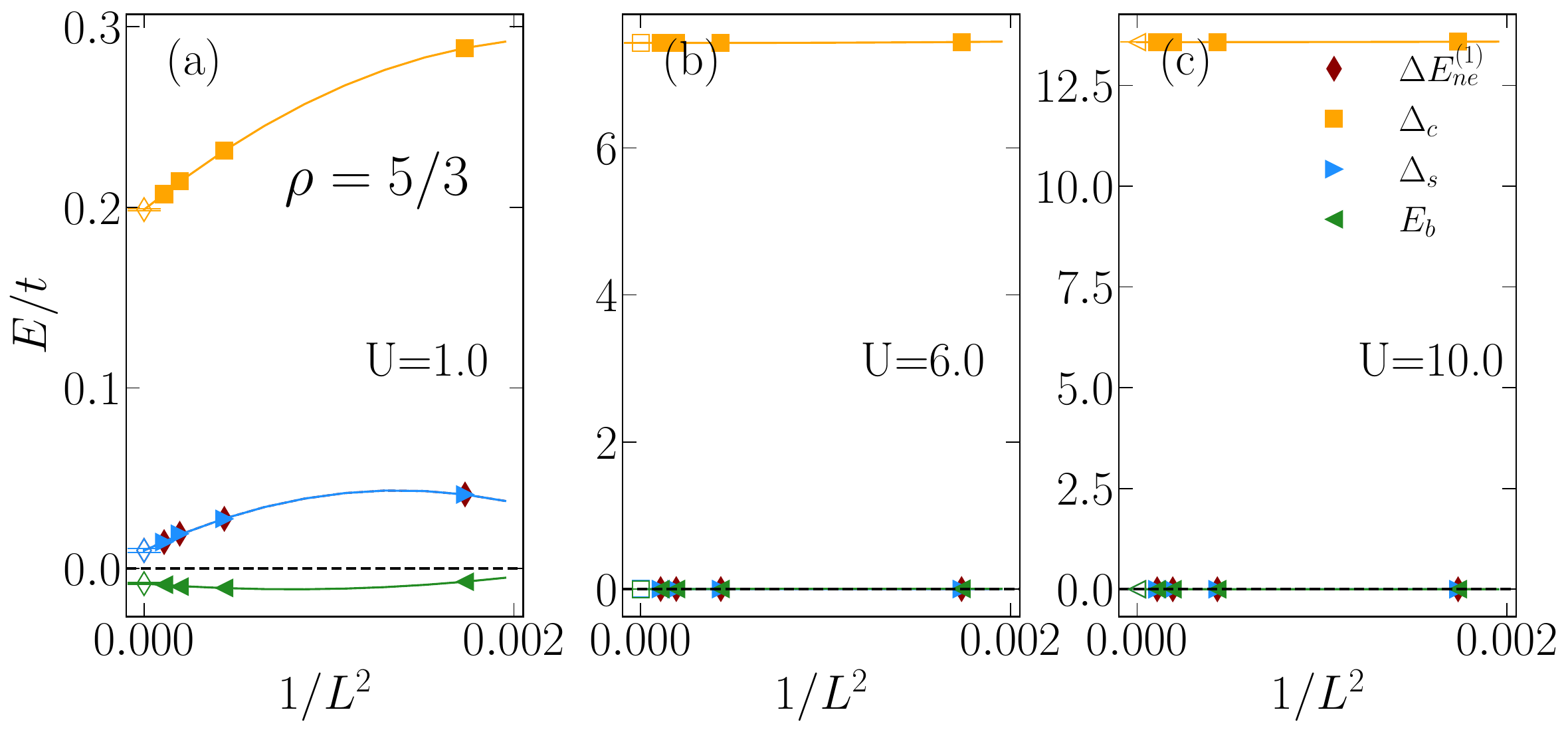}
    \caption{Finite-size scaling of the thermodynamic energy gaps at filling $\rho = 5/3$. The energy gaps are plotted versus $1/L^2$ for (a) $U=1.0$, (b) $U=6.0$ and (c) $U=10.0$. Solid symbols denote the exact DMRG data for finite-size systems ($L = 24, 48, 72$, and $96$ under PBC). Open symbols denote the extrapolated values in the thermodynamic limit ($L \to \infty$).}
    \label{fig:fitting_rho5_3}
\end{figure}

For completeness, we also present the scaling results for the reference fillings $\rho = 1$ and $\rho = 5/3$ in Figs.~\ref{fig:fitting_rho1} and \ref{fig:fitting_rho5_3}. These fillings provide useful comparisons with more conventional correlated insulating regimes. In the parameter ranges shown, the interaction opens a finite charge gap, while the spin gap extrapolates to zero or remains much smaller than the charge gap in the thermodynamic limit. These reference cases therefore differ from the $\rho=2/3$ correlated pump, where both charge and spin gaps are finite in the bulk regime relevant to the interaction-split boundary spectral flow.

The scaling results for $\rho=4/3$ in Fig.~\ref{fig:fitting_rho4_3} are included as a filling-dependent comparison. At weak interaction [Fig.~\ref{fig:fitting_rho4_3}(a)], the system is fully gapped and belongs to the $C=-2$ pump sector. Near the interaction-driven transition [Fig.~\ref{fig:fitting_rho4_3}(b)], the neutral and spin gaps are strongly suppressed, while the charge gap remains finite within the finite-size scaling. At stronger interaction [Fig.~\ref{fig:fitting_rho4_3}(c)], the system enters a different gapped regime. These results illustrate that spatially modulated interactions can produce filling-dependent topological responses. however, in the main text we use this filling only as a comparison. The central boundary-fractionalization mechanism is established at $\rho=2/3$.

\begin{figure}[tb]
    \centering
    \includegraphics[width=1\columnwidth]{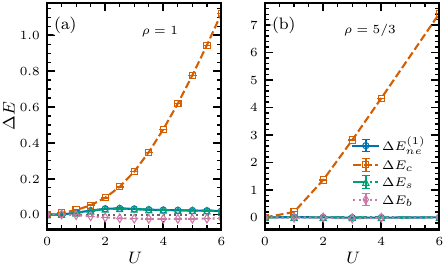}
    \caption{Extrapolated thermodynamic energy gaps as a function of the Hubbard interaction strength $U$ for the reference fillings (a) $\rho = 1$ and (b) $\rho = 5/3$. The plotted quantities include the bulk charge gap $\Delta_c$, spin gap $\Delta_s$, lowest neutral excitation gap $\Delta E_{ne}^{(1)}$, and pair-binding energy $E_b$. At both fillings, the system develops a finite charge gap while the spin excitations remain gapless or nearly gapless in the thermodynamic limit, distinguishing these reference regimes from the fully gapped $\rho=2/3$ correlated pump discussed in the main text.}
    \label{fig:gaps_trivial}
\end{figure}

To further contrast the $\rho=2/3$ pump with reference correlated phases, Fig.~\ref{fig:gaps_trivial} presents the extrapolated thermodynamic gaps for $\rho=1$ and $\rho=5/3$. At $\rho=1$, the interaction drives a charge-gapped Mott regime with a spin sector that remains gapless in the thermodynamic limit, consistent with the expected low-energy magnetic physics of a one-dimensional Mott insulator. A similar qualitative behavior is found at $\rho=5/3$, where the modulated interaction opens a finite charge gap but does not produce the same finite spin gap structure observed at $\rho=2/3$. These results show that the fully gapped correlated pump at $\rho=2/3$ is distinct from the reference Mott regimes and provides the appropriate bulk setting for the interaction-split boundary spectral flow discussed in the main text.

\section{S5. MANY-BODY CHERN NUMBER}
\label{sec:Chern}

For noninteracting pumps, the Chern number can be computed from the occupied single-particle Bloch bands. In the interacting Hubbard chain studied here, we instead use a many-body formulation based on twisted boundary conditions~\cite{Niu1985,Xiao2010}. For the total charge pump discussed in the main text, we impose the same twist angle on both spin species, so that a fermion hopping across the boundary acquires the phase $\theta$:
\begin{equation}
    \hat{c}_{L+1,\sigma} = e^{i\theta} \hat{c}_{1,\sigma}.
    \label{eq:TBC}
\end{equation}

The many-body ground state $|\Psi_0(\theta,\phi)\rangle$ is then defined over the two-dimensional parameter space spanned by the twist angle $\theta \in [0, 2\pi]$ and the pump phase $\phi \in [0, 2\pi]$. The corresponding many-body Chern number is
\begin{equation}
    C = \frac{1}{2\pi} \int_{0}^{2\pi} d\theta \int_{0}^{2\pi} d\phi \left( \partial_\theta A_\phi - \partial_\phi A_\theta \right),
\end{equation}
where $A_\mu = i \langle \Psi_0(\theta,\phi) | \partial_\mu | \Psi_0(\theta,\phi) \rangle$ is the many-body Berry connection. This invariant gives the quantized charge transported over one adiabatic cycle, provided that the many-body gap remains open over the $(\theta,\phi)$ torus.

In the numerical calculations, we evaluate this invariant using the gauge-invariant lattice discretization of the Berry curvature~\cite{Fukui2005,Varney2011}. The parameter space is divided into a discrete grid of $(N_\theta, N_\phi) = (16, 16)$ points. At each grid point $\mathbf{q} = (\theta, \phi)$, we define the U(1) link variables
\begin{equation}
    \mathcal{U}_\mu(\mathbf{q}) = \frac{\langle \Psi_0(\mathbf{q}) | \Psi_0(\mathbf{q} + \Delta\hat{\mu}) \rangle}{|\langle \Psi_0(\mathbf{q}) | \Psi_0(\mathbf{q} + \Delta\hat{\mu}) \rangle|}.
\end{equation}

The Berry flux through each elementary plaquette is obtained from the product of link variables around that plaquette, and the many-body Chern number is obtained by summing the flux over the full discretized parameter space. With the grid size used here, the resulting Chern numbers are quantized to integers within numerical precision. At filling $\rho=2/3$, this procedure gives $C=+2$ throughout the gapped regime discussed in the main text, confirming that the interaction-split boundary spectral flow occurs within an integer charge pump.

\section{S6. Hartree Reference Pump and Its Limitations}
\label{sec:hartree}

To distinguish the single-particle Hartree contribution from genuine Hubbard-correlation effects, we compare the full interaction-modulated model with its Hartree mean-field reference. This comparison is useful because the sliding interaction modulation acts at finite density and therefore naturally generates an effective scalar potential even before correlation effects are included.

We begin with the full onsite Hubbard interaction, $H_{\text{int}} = \sum_i U_i(\phi) \hat{n}_{i,\uparrow} \hat{n}_{i,\downarrow}$. In the weak-coupling regime, we can decompose the number operators into their expectation values and quantum fluctuations: $\hat{n}_{i,\sigma} = \langle \hat{n}_{i,\sigma} \rangle + \delta\hat{n}_{i,\sigma}$. Neglecting second-order fluctuations ($\delta\hat{n}_{i,\uparrow} \delta\hat{n}_{i,\downarrow} \approx 0$) yields the standard Hartree mean-field decoupling:
\begin{equation}
    \hat{n}_{i,\uparrow} \hat{n}_{i,\downarrow} \approx \langle \hat{n}_{i,\uparrow} \rangle \hat{n}_{i,\downarrow} + \langle \hat{n}_{i,\downarrow} \rangle \hat{n}_{i,\uparrow} - \langle \hat{n}_{i,\uparrow} \rangle \langle \hat{n}_{i,\downarrow} \rangle.
\end{equation}
For an unpolarized state preserving global SU(2) symmetry, the average density per spin component is $\langle \hat{n}_{i,\uparrow} \rangle = \langle \hat{n}_{i,\downarrow} \rangle = \rho/2$ at the level of this spatially averaged Hartree approximation. Substituting this into the interaction term gives the noninteracting Hartree reference Hamiltonian:
\begin{equation}
    H_{\text{H}} = -t \sum_{i,\sigma} \left( \hat{c}^\dagger_{i,\sigma} \hat{c}_{i+1,\sigma} + \text{H.c.} \right) + \sum_i V_i^{\text{H}}(\phi) \hat{n}_i,
\end{equation}
where $V_i^{\text{H}}(\phi) = \frac{\rho}{2} U_i(\phi)$ acts as an effective, spin-independent sliding scalar potential. 

In this Hartree reference, the Hamiltonian separates into two identical noninteracting spin sectors: $H_{\text{H}} = H_{\uparrow} \oplus H_{\downarrow}$. At $\rho=2/3$, the period-three sliding potential $V_i^{\text{H}}(\phi)$ opens a single-particle band gap and realizes a commensurate Harper pump in each spin channel. Since each independent spin sector carries Chern number $C_{\sigma}=1$, the Hartree model captures the same net integer pumped charge, $Q_{\text{pump}}=C_{\uparrow}+C_{\downarrow}=2$, as the interacting pump discussed in the main text.

The important difference is not the total pumped charge, but how this integer pump is implemented at an open boundary. In the Hartree reference model, $H_{\uparrow}$ and $H_{\downarrow}$ are identical and uncoupled. Their edge states are therefore spin-degenerate, so the spin-up and spin-down boundary modes cross the gap at the same pump phase. This gives the simultaneous edge spectral flow sketched in Fig. 1(b) of the main text.

The full Hubbard model has the same integer bulk response at $\rho=2/3$, but the open-boundary spectrum is reconstructed by local correlations. In particular, the boundary spectral flow is split into two separated edge events, and the intermediate window supports a low-energy neutral spin excitation while charge addition and removal remain gapped. These features are absent in the Hartree reference pump. The comparison therefore shows that the Hartree channel accounts for the existence of a $C=2$ integer pump, but it does not capture the interaction-fractionalized boundary spectral flow.

The bulk excitation gaps provide a complementary distinction. In a spin-degenerate Hartree band pump, charge and spin excitations are both controlled by the same underlying single-particle band gap, and no independent low-energy neutral spin sector is generated. By contrast, the DMRG results in Fig. 4 of the main text show separated charge and spin energy scales in the correlated pump, with $\Delta_s<\Delta_c$ throughout the gapped $\rho=2/3$ regime studied. This separation supports the interpretation that the low-energy neutral excitation is predominantly spin-like, while charged excitations remain gapped. Thus, although the Hartree reference correctly captures the net integer charge pump, it is insufficient to describe the correlated boundary implementation of that pump in the full Hubbard model.

\section{S7. Spin-Resolved Chern Numbers and Berry-Flux Distributions}
\label{sec:spinChern}

To further characterize the bulk topological response, we evaluate many-body Chern numbers under spin-resolved twisted boundary conditions. We impose
\[
\theta_\uparrow=m_\uparrow\theta,\qquad
\theta_\downarrow=m_\downarrow\theta,
\]
where $m_\uparrow,m_\downarrow=0,\pm1$ and $\theta\in[0,2\pi]$ is the twist angle used in the Chern-number calculation. A uniform charge twist corresponds to $(m_\uparrow,m_\downarrow)=(1,1)$, while a spin-selective twist corresponds to $(m_\uparrow,m_\downarrow)=(1,0)$ or $(0,1)$. This provides a useful diagnostic for whether the bulk charge pumping response is additive in the two spin components.

\begin{table}[tb]
\centering
\caption{Many-body Chern numbers at filling $\rho=2/3$ under spin-resolved twisted boundary conditions. The row and column labels denote the twist coefficients $m_\uparrow$ and $m_\downarrow$, defined by $\theta_\uparrow=m_\uparrow\theta$ and $\theta_\downarrow=m_\downarrow\theta$.}
\label{tab:twC_2d3}
\renewcommand{\arraystretch}{1.3}
\setlength{\tabcolsep}{10pt}
\begin{tabular}{c r r r}
\toprule
\multirow{2}{*}{$m_{\uparrow}$} & \multicolumn{3}{c}{$m_{\downarrow}$} \\
\cmidrule(lr){2-4}
& $+1$ & $0$ & $-1$ \\
\midrule
$+1$ & 2 & 1 & 0 \\
$0$  & 1 & $\times$ & $-1$ \\
$-1$ & 0 & $-1$ & $-2$ \\
\bottomrule
\end{tabular}
\end{table}

\begin{figure}[tb]
\centering
\includegraphics[width=0.95\columnwidth]{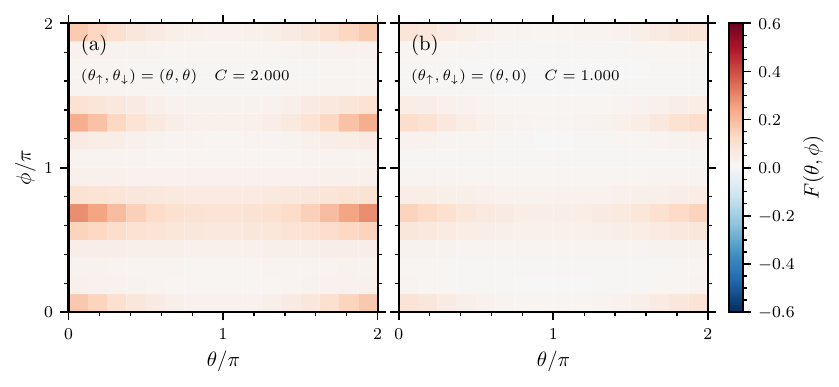}
    \caption{Berry-flux distributions at $\rho=2/3$ for $L=24$, $U=2.0$, and $\delta=1.0$.
(a) Uniform charge twist, $(\theta_\uparrow,\theta_\downarrow)=(\theta,\theta)$, giving $C=2$.
(b) Spin-selective twist, $(\theta_\uparrow,\theta_\downarrow)=(\theta,0)$, giving $C=1$.
The two panels are plotted using the same color scale.
The Berry-flux distribution in (a) is approximately twice that in (b), consistent with an additive bulk response from two spin components.}
\label{fig:berryF_rho2_3}
\end{figure}

Table~\ref{tab:twC_2d3} summarizes the spin-resolved responses for the correlated insulating phase at filling $\rho=2/3$. A uniform charge twist gives the total Chern number $C=2$, corresponding to the pumping of two units of charge per cycle. A spin-selective twist applied only to one spin component gives $C=1$, while opposite twists applied to the two spin components give zero net charge response. These results are additive in the sense that the uniform charge response equals the sum of the two spin-selective responses.

The corresponding Berry-flux distributions are shown in Fig.~\ref{fig:berryF_rho2_3}. The flux distribution for the uniform twist, $(\theta_\uparrow,\theta_\downarrow)=(\theta,\theta)$, is approximately twice that for the spin-selective twist, $(\theta_\uparrow,\theta_\downarrow)=(\theta,0)$, not only after integration over the twist-pump torus but also at the level of the local Berry flux. This supports the interpretation that, at $\rho=2/3$, the bulk response remains equivalent to two additive $C=1$ spin components. Therefore the main effect of the Hubbard interaction in the $\rho=2/3$ pump is not a fractionalization of the bulk pumped charge. Rather, as shown in the main text, local correlations reconstruct the open-boundary implementation of this integer pump by splitting the boundary spectral flow and producing an intermediate neutral spinful boundary sector.

\begin{table}[tb]
\centering
\caption{Many-body Chern numbers at filling $\rho=4/3$ for $U<U_c$. The labels $m_\uparrow$ and $m_\downarrow$ are defined by $\theta_\uparrow=m_\uparrow\theta$ and $\theta_\downarrow=m_\downarrow\theta$.}
\label{tab:twC_4d3_smallU}
\renewcommand{\arraystretch}{1.3}
\setlength{\tabcolsep}{10pt}
\begin{tabular}{c r r r}
\toprule
\multirow{2}{*}{$m_{\uparrow}$} & \multicolumn{3}{c}{$m_{\downarrow}$} \\
\cmidrule(lr){2-4}
& $+1$ & $0$ & $-1$ \\
\midrule
$+1$ & $-2$ & $-1$ & $0$ \\
$0$  & $-1$ & $\times$ & $1$ \\
$-1$ & $0$ & $1$ & $2$ \\
\bottomrule
\end{tabular}
\end{table}

For comparison, Table~\ref{tab:twC_4d3_smallU} shows the corresponding spin-resolved responses at filling $\rho=4/3$ in the weak-interaction regime, $U<U_c$. In this regime, the response is also additive, but with the opposite sign compared with $\rho=2/3$: a uniform charge twist gives $C=-2$, while a spin-selective twist gives $C=-1$. This behavior is consistent with a weak-coupling pump whose bulk response can still be decomposed into two spin components.

\begin{table}[tb]
\centering
\caption{Many-body Chern numbers at filling $\rho=4/3$ for $U>U_c$. The labels $m_\uparrow$ and $m_\downarrow$ are defined by $\theta_\uparrow=m_\uparrow\theta$ and $\theta_\downarrow=m_\downarrow\theta$.}
\label{tab:twC_4d3_largeU}
\renewcommand{\arraystretch}{1.3}
\setlength{\tabcolsep}{10pt}
\begin{tabular}{c r r r}
\toprule
\multirow{2}{*}{$m_{\uparrow}$} & \multicolumn{3}{c}{$m_{\downarrow}$} \\
\cmidrule(lr){2-4}
& $+1$ & $0$ & $-1$ \\
\midrule
$+1$ & 1 & 2 & 0 \\
$0$  & 2 & $\times$ & $-2$ \\
$-1$ & 0 & $-2$ & $-1$ \\
\bottomrule
\end{tabular}
\end{table}

\begin{figure}[tb]
\centering
\includegraphics[width=0.95\columnwidth]{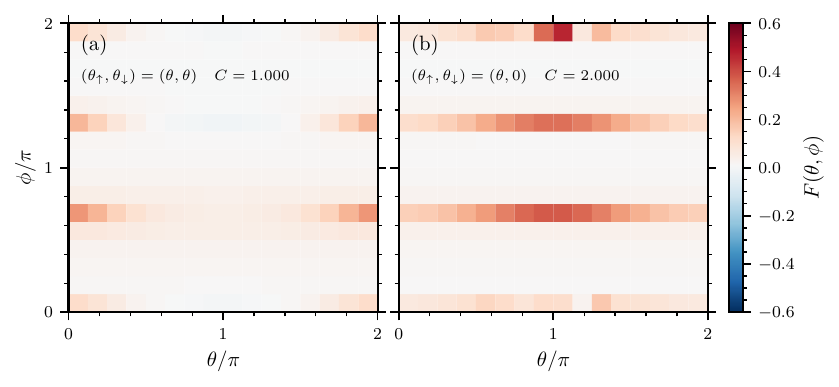}
\caption{
Berry-flux distributions at $\rho=4/3$ in the strong-interaction regime, shown for $L=24$, $U=8.0$, and $\delta=0.1$.
(a) Uniform charge twist, $(\theta_\uparrow,\theta_\downarrow)=(\theta,\theta)$, giving $C=1$.
(b) Spin-selective twist, $(\theta_\uparrow,\theta_\downarrow)=(\theta,0)$, giving $C=2$.
The two panels are plotted using the same color scale.
Unlike the $\rho=2/3$ case, the uniform-twist flux distribution is not twice the spin-selective one, reflecting a non-additive spin-resolved bulk response.
}
\label{fig:berryF_rho4_3}
\end{figure}

At $\rho=4/3$ and stronger interaction, $U>U_c$, the spin-resolved response changes qualitatively, as shown in Table~\ref{tab:twC_4d3_largeU}. The uniform charge twist gives a total Chern number $C=1$, whereas a spin-selective twist gives $C=2$. Thus the response is no longer additive in the simple spin-channel sense. This non-additivity is also visible directly in Fig.~\ref{fig:berryF_rho4_3}: the Berry-flux distribution for $(\theta_\uparrow,\theta_\downarrow)=(\theta,\theta)$ is not related to the spin-selective distribution by a simple factor of two.

This breakdown of additivity indicates that, at $\rho=4/3$ and large $U$, twisting one spin component probes a correlated many-body response rather than an independent single-spin channel. A useful physical interpretation is that the finite doublon density becomes important in this regime. Because the motion of one spin species is strongly constrained by the other through the on-site interaction, a spin-selective twist can induce a collective response involving both spin components. The resulting Chern numbers therefore need not follow the additive pattern observed at $\rho=2/3$.

The $\rho=4/3$ results are included as a filling-dependent comparison. They show that spatially modulated Hubbard interactions can generate strongly correlated, non-additive pumping responses at other commensurate fillings. The main result of the Letter, however, concerns the $\rho=2/3$ phase, where the bulk Chern response remains additive and integer-valued, while the open-boundary spectral flow is reconstructed by local correlations.

\section{S8. Additional Edge Spectra at $\rho=4/3$}
\label{sec:rho4_3_edge}

\begin{figure}[tb]
    \centering
    \includegraphics[width=0.95\columnwidth]{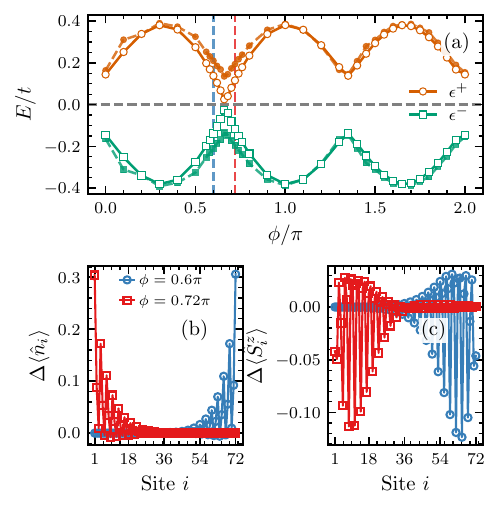} 
    \caption{Single-particle excitation spectrum and real-space fermionic edge response at filling $\rho=4/3$. (a) Addition and removal energies ($\epsilon^\pm$) versus $\phi$ for $L=72$ at $U=8.0$ and $\delta=0.1$. Filled and open markers denote PBC and OBC results, respectively. The PBC spectrum remains gapped, while OBC edge modes cross the gap near $\phi \approx 2\pi/3$. Real-space heatmaps of (b) $\Delta \langle \hat{n}_i \rangle$ and (c) $\Delta \langle \hat{S}^z_i \rangle$ around this crossing show charge and spin responses localized near the boundary.}
    \label{fig:fermion_spec_rho4_3}
\end{figure}

\begin{figure}[tb]
    \centering
    \includegraphics[width=0.95\columnwidth]{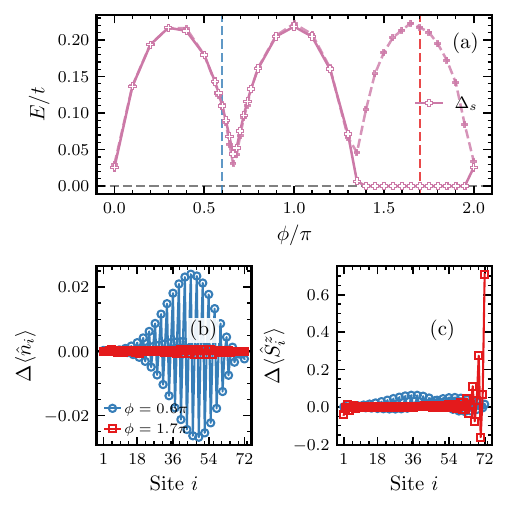} 
    \caption{Spin-flip excitation spectrum and real-space neutral spin edge response at filling $\rho=4/3$. (a) Lowest neutral spin-flip excitation energy versus $\phi$ for $L=72$ at $U=8.0$ and $\delta=0.1$. Filled and open markers denote PBC and OBC results, respectively. The PBC spin gap remains finite, while the OBC spin-flip excitation is strongly suppressed near $\phi \approx 5\pi/3$. Real-space heatmaps of (b) $\Delta \langle \hat{n}_i \rangle$ and (c) $\Delta \langle \hat{S}^z_i \rangle$ show a localized boundary spin response with little accompanying charge response.}
    \label{fig:magnon_spec_rho4_3}
\end{figure}

In this section we provide additional open-boundary spectra for the large-$U$ regime at filling $\rho=4/3$. These results serve as a filling-dependent comparison to the $\rho=2/3$ integer pump discussed in the main text. They are not used to establish the interaction-fractionalized boundary spectral flow at $\rho=2/3$, but illustrate that the same interaction modulation can generate distinct boundary responses at other commensurate fillings.

As the pump phase $\phi$ evolves, the open-chain spectrum first exhibits a fermionic boundary crossing. Figure~\ref{fig:fermion_spec_rho4_3}(a) shows the single-particle addition and removal energies at $U=8.0$. Under periodic boundary conditions, the spectrum remains gapped, consistent with an insulating bulk. Under open boundary conditions, boundary modes enter the bulk gap near $\phi \approx 2\pi/3$. The corresponding real-space profiles in Figs.~\ref{fig:fermion_spec_rho4_3}(b) and \ref{fig:fermion_spec_rho4_3}(c) show that both the local charge response $\Delta \langle \hat{n}_i \rangle$ and the spin response $\Delta \langle \hat{S}^z_i \rangle$ are concentrated near the boundary during this event.

At a later pump phase, near $\phi \approx 5\pi/3$, the low-energy boundary response appears in the neutral spin sector. As shown in Fig.~\ref{fig:magnon_spec_rho4_3}(a), the lowest open-chain spin-flip excitation is strongly suppressed while the corresponding periodic-chain spin gap remains finite. The real-space profiles in Figs.~\ref{fig:magnon_spec_rho4_3}(b) and \ref{fig:magnon_spec_rho4_3}(c) show that this low-energy excitation has a localized spin response near the edge, with little accompanying charge response. We therefore refer to this feature as a neutral spin edge response.

Taken together, Figs.~\ref{fig:fermion_spec_rho4_3} and \ref{fig:magnon_spec_rho4_3} show that, at $\rho=4/3$ and large $U$, charged and neutral spin boundary responses occur at different pump phases. This behavior is consistent with the non-additive spin-resolved Chern response discussed in Sec.~\ref{sec:spinChern}. It provides an additional example of filling-dependent boundary reconstruction induced by spatially modulated interactions. However, the main result of the Letter, remains the $\rho=2/3$ integer $C=+2$ pump, where the total bulk charge response is additive while the open-boundary spectral flow separates into charged and neutral spin sectors.


\end{document}